\documentclass[11pt]{article}
\usepackage[T1]{fontenc}
\usepackage[utf8]{inputenc}
\usepackage{lmodern,microtype}
\usepackage[margin=1in]{geometry}
\usepackage{amsmath,amssymb,bm,booktabs,tabularx,graphicx,xcolor}
\usepackage[numbers,sort&compress]{natbib}
\usepackage[colorlinks=true,linkcolor=blue!45!black,citecolor=blue!45!black,urlcolor=blue!45!black]{hyperref}
\usepackage{enumitem}
\setlist{nosep}
\newcommand{\SF}{\mathrm{SF}}
\newcommand{\NSF}{\mathrm{NSF}}
\newcommand{\Q}{\bm{Q}}
\title{Resolving magnetic order and dynamics with polarized neutron scattering: a practical guide to longitudinal polarization analysis}
\author{Sijie Xu$^{1,2,*}$ and Pengcheng Dai$^{1,2,\dagger}$\\[0.5em]
\small $^{1}$Department of Physics and Astronomy, Rice University,\\
\small Houston, Texas 77005, USA\\[0.3em]
\small $^{2}$Rice Laboratory for Emergent Magnetic Materials and Smalley-Curl Institute,\\
\small Rice University, Houston, Texas 77005, USA}
\hypersetup{pdfauthor={Sijie Xu and Pengcheng Dai}}
\date{October 2026}
\begin{document}
\maketitle\vspace{-1.5em}
\begingroup
\renewcommand{\thefootnote}{\ensuremath{*}}
\footnotetext[1]{Email: \href{mailto:sx27@rice.edu}{sx27@rice.edu}}
\renewcommand{\thefootnote}{\ensuremath{\dagger}}
\footnotetext[2]{Email: \href{mailto:pdai@rice.edu}{pdai@rice.edu}}
\endgroup

\begin{abstract}
Neutron scattering provides microscopic information on magnetic order and spin dynamics. Most experiments, however, use unpolarized neutrons and do not fully exploit the information carried by the neutron spin. Longitudinal polarization analysis (LPA), established in the pioneering work of Moon, Riste, and Koehler at Oak Ridge National Laboratory in the late 1960s, separates magnetic and nonmagnetic scattering contributions through their different effects on the neutron spin. In this review, we illustrate how LPA can distinguish weak magnetic scattering from nonmagnetic contributions, constrain small moment components, and resolve the spin-space anisotropy of magnetic excitations. We introduce the formalism connecting measured polarization channels to magnetic quantities and discuss scattering geometry, crystallographic projections, and corrections for finite polarization. Our aim is to provide a practical guide for neutron scatterers using LPA to determine magnetic structures and characterize spin excitations in quantum materials. We conclude with opportunities in altermagnets, unconventional superconductors, and kagome magnets.

\end{abstract}

\section{Introduction}
Neutron scattering has played a central role in establishing our microscopic understanding of magnetic order and spin dynamics. Neutrons are electrically neutral, carry a magnetic moment, and penetrate deeply into many bulk materials. Cold and thermal neutrons have wavelengths and energies well matched to atomic structures and magnetic excitations. These properties enable neutron diffraction to determine magnetic structures and inelastic scattering to map spin dynamics in momentum (\(\bm{q}\)) and energy space \(E = \hbar\omega\) \cite{Zaliznyak2005,Boothroyd2020}.

Most experiments use unpolarized neutron beams. Their magnetic contribution is the summed response perpendicular to the scattering wave vector \(\Q = \bm{k}_{\mathrm{i}} - \bm{k}_{\mathrm{f}}\), where \(\bm{k}_{\mathrm{i}}\) and \(\bm{k}_{\mathrm{f}}\) are the wave vectors of the incident and final neutron beams, respectively. For many problems, this information is sufficient. However, some questions require information that is absent from, or poorly constrained by, the unpolarized intensity alone. A weak magnetic contribution may be difficult to distinguish from nonmagnetic backgrounds; different moment configurations may produce similar intensities; and overlapping excitations may involve several spin components. Wave-vector (\(\Q\)) and temperature dependence provide useful constraints but do not, by themselves, always distinguish these possibilities unambiguously \cite{Zaliznyak2005,Boothroyd2020}.

Neutron polarization analysis provides this information through spin-dependent selection rules. Controlling the incident polarization and analyzing the scattered polarization separates processes according to their effects on the neutron spin. In LPA, the incident and analyzed spin quantization axes are parallel or antiparallel to the same chosen axis. The neutron spin-flip (SF) and non-spin-flip (NSF) cross sections then give complementary projections of the magnetic and nonmagnetic responses. The classic experiments of Moon, Riste, and Koehler at the High Flux Isotope Reactor established this approach \cite{101103PhysRev181920}.

In addition, comparing cross sections measured with opposite incident polarization provides access to additional terms. For neutron polarization parallel to \(\Q\), the difference between the \(+ -\) (incident neutron spin along \(\Q\) being \(+\) or up spin, and scattered neutron spin antiparallel being \(-\) or down spin) and \(- +\) SF cross sections measures a chiral magnetic contribution \cite{PhysRevB.75.054418}. For neutron polarization perpendicular to \(\Q\), differences between the \(+ +\) and \(- -\) NSF cross sections can measure nuclear-magnetic interference. These are distinct observables: comparing the two SF channels does not, in this geometry, measure nuclear-magnetic interference. Spherical neutron polarimetry, implemented with devices such as Cryopad \cite{TASSET199969}, also measures transverse polarization transfer and provides further constraints on complex magnetic structures. Polarization analysis can therefore determine magnetic handedness, constrain magnetization distributions, and, at suitable reflections, distinguish time-reversed magnetic domain populations \cite{101103PhysRev1301670,Maleev1963,McClarty2025}. Their interpretation requires careful treatment of geometry, polarization efficiency, and symmetry.

In this review, we use selected studies, drawn primarily from our group's work, to illustrate how LPA can provide information not readily accessible from unpolarized measurements alone. Rather than attempting a comprehensive survey, we organize the discussion around several recurring experimental questions: how to identify weak magnetic scattering, how to detect or constrain small additional magnetic components, and how to use neutron polarization analysis to determine the spin-space anisotropy of magnetic excitations. The examples include studies of quantum spin ice, anomalous transport, and uniaxial strain-dependent magnetism, and are chosen to demonstrate how polarization analysis can be applied in practice to address different physical questions. Section~\ref{sec:principles} introduces the physical basis of the measurements and the practical considerations needed for their interpretation. Through this discussion, we aim to provide a practical guide for neutron scatterers interested in applying polarization analysis to unveil unique magnetic properties in quantum materials. We close by highlighting emerging opportunities for polarized neutron scattering in several active areas of quantum materials research, including altermagnets, unconventional superconductors, and kagome magnets.

\section{Magnetic information from polarization analysis}
\label{sec:principles}

Polarization analysis separates magnetic scattering according to the neutron spin and thereby reveals information beyond the summed scattering intensity. The key ingredients for extracting such information are the local polarization frame, the corresponding neutron spin-flip (SF) and non-spin-flip (NSF) selection rules, the relation between local and crystallographic spin components, and the finite neutron polarization efficiency of a real instrument. Throughout the paper, $\Q=\bm{k}_{\mathrm{i}}-\bm{k}_{\mathrm{f}}$ denotes the scattering wave vector, conventionally called momentum transfer. $\Q=\bm{q}+\bm{G}$, where $\bm{q}$ is momentum within the first Brillouin zone, and $\bm{G}$ is a reciprocal-lattice vector. In a primitive reciprocal basis, $\bm{G}=h\bm{a}^\ast+k\bm{b}^\ast+l\bm{c}^\ast$ with integer $h$, $k$, $l$. A reciprocal-lattice point need not produce an observable nuclear Bragg peak: its nuclear structure factor may vanish. The neutron energy transfer to the sample is $E=\hbar\omega=E_{\mathrm{i}}-E_{\mathrm{f}}$.

\subsection{Separating local magnetic responses}
\label{sec:crosssections}

\begin{figure}[htb]
\centering
\includegraphics[width=0.85\linewidth]{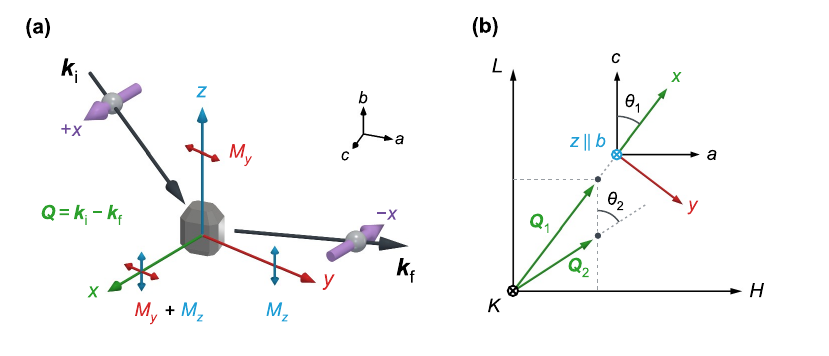}
\caption{\textbf{Scattering geometry and magnetic selection rules for longitudinal polarization analysis.}
\textbf{(a)} The incident and scattered neutron wave vectors define $\Q=\bm{k}_{\mathrm{i}}-\bm{k}_{\mathrm{f}}$. The local polarization axes are $x\parallel\Q$, $z$ normal to the scattering plane, and $y$ chosen to complete a right-handed coordinate system. Violet arrows illustrate reversal of the neutron spin from $+x$ to $-x$ in a spin-flip (SF) scattering event. Red and blue double-headed arrows indicate the magnetic components contributing to $M_y$ and $M_z$, respectively. The labels $M_y+M_z$, $M_z$, and $M_y$ indicate the magnetic contributions to the SF channels for neutron polarizations along $x$, $y$, and $z$.
\textbf{(b)} Relation between the local polarization frame and the crystallographic axes in the $(H0L)$ scattering plane. For different momentum-transfer vectors $\Q_1$ and $\Q_2$, the local $x$ axis follows $\Q$ and the local $y$ axis rotates within the $ac$ plane, while $z\parallel b$ remains fixed. Measurements at different wave vectors therefore provide different projections of the same magnetic response.}
\label{fig:lpa_geometry}
\end{figure}

The starting point of LPA is the local coordinate system defined by the scattering geometry. In the conventional LPA frame, incident neutron polarization directions are $x$, $y$, and $z$, where 
$x\parallel\Q$, $z$ is normal to the scattering plane, and $y$ is chosen to complete a right-handed coordinate system [Fig.~\ref{fig:lpa_geometry}(a)]. Magnetic neutron scattering probes only magnetization components perpendicular to $\Q$. Polarization analysis then separates this transverse response according to its orientation relative to the neutron spin: components perpendicular to the neutron polarization contribute to neutron SF scattering, whereas parallel components contribute to NSF scattering~\cite{101103PhysRev181920,101103RevModPhys87855}.

We denote by $M_\alpha$ the magnetic scattering contribution associated with magnetization polarized along direction $\alpha$ at the measured momentum and energy $(\Q,E)$ \cite{Boothroyd2020}. For elastic scattering, $M_\alpha$ is proportional to the squared component of the magnetic structure factor perpendicular to $\Q$, resolved along the local direction $\alpha$, and is therefore related to the square of the corresponding moment projection through the magnetic structure factor. For inelastic scattering, it denotes the corresponding component of the dynamical magnetic response. Thus, $M_\alpha$ is a scattering intensity rather than a magnetic moment.\footnote{The magnetic form-factor dependence is retained in $M_\alpha$; it has not been divided out.}

For ideal polarization, and momentarily leaving aside nuclear spin-incoherent scattering, background, chirality, and nuclear--magnetic interference, the neutron SF cross sections read
\begin{equation}
\sigma_x^{\SF}\propto M_y+M_z,\qquad
\sigma_y^{\SF}\propto M_z,\qquad
\sigma_z^{\SF}\propto M_y.
\label{eq:ideal_sf}
\end{equation}

For elastic scattering (neutron diffraction), the consequence of these selection rules is apparent. The quantities $M_y$ and $M_z$ can be calculated directly from a proposed magnetic structure. We define the vector magnetic structure factor before projection perpendicular to $\Q$ as~\cite{Zaliznyak2005,ShiraneShapiroTranquada2002}
\begin{equation}
\bm{F}_{\mathrm{M}}(\Q)=
\frac{|\gamma| r_0}{2}
\sum_j
f_j(Q)\mathrm{e}^{-W_j(\Q)}
\bm{\mu}_j
\mathrm{e}^{\mathrm{i}\Q\cdot\bm{r}_j},
\label{eq:mag_structure_factor}
\end{equation}
where $|\gamma| r_0/2=2.695~\mathrm{fm}$ is the magnetic scattering-length constant, $f_j(Q)$ ($Q=|\Q|$) is the magnetic form factor of the magnetic ion at site $j$, $\mathrm{e}^{-W_j(\Q)}$ is its Debye--Waller factor, and $\bm{\mu}_j$ is the magnetic moment at site $j$, expressed in units of $\mu_{\mathrm{B}}$. Since $\hat{\bm{y}}$ and $\hat{\bm{z}}$ are both perpendicular to $\Q$, the two polarization-resolved magnetic intensities are
\begin{equation}
M_y(\Q)=
|\bm{F}_{\mathrm{M}}\cdot\hat{\bm{y}}|^2,
\qquad
M_z(\Q)=
|\bm{F}_{\mathrm{M}}\cdot\hat{\bm{z}}|^2.
\label{eq:elastic_MyMz}
\end{equation}

Thus, once the moment configuration and scattering wave vector are specified, the expected $M_y$ and $M_z$ follow from simple projections in the local frame and can be compared directly with experiment. By contrast, an unpolarized measurement constrains only their sum.

The complete ideal polarized-neutron cross sections are summarized in Table~\ref{tab:crosssections}. In addition to $M_y$ and $M_z$, they contain contributions from nuclear coherent scattering, nuclear spin-incoherent scattering, chiral magnetic correlations, and nuclear--magnetic interference. Differences between cross sections measured with reversed neutron spin states can isolate chiral or nuclear--magnetic interference contributions. Averaging the two incident-spin states within each SF or NSF channel cancels the terms that are odd under incident-spin reversal. Both operations require the measurements to be corrected to the same efficiency and normalization. Here, reversing the incident spin refers to the neutron polarization, not the sample's magnetic domain. Chiral magnetic correlations measured in this way should also be distinguished from scalar three-spin chirality, which cannot in general be inferred directly from a SF asymmetry alone. Detailed expressions and applications of these terms are discussed extensively in standard treatments of polarized-neutron scattering~\cite{101103PhysRev181920,ShiraneShapiroTranquada2002} and will not be reproduced comprehensively here.

\begin{table}[htbp]
\centering
\small
\renewcommand{\arraystretch}{1.5}
\begin{tabular}{@{}lll@{}}
\toprule
Polarization axis & NSF: $\sigma_\alpha^{\pm\pm}$ & SF: $\sigma_\alpha^{\pm\mp}$ \\
\midrule
$x\parallel\Q$ & $N+\tfrac13 I_{\mathrm{s}}+B$ &
$M_y+M_z\pm C_x+\tfrac23 I_{\mathrm{s}}+B$ \\
$y$ & $N+M_y\pm I_y+\tfrac13 I_{\mathrm{s}}+B$ &
$M_z+\tfrac23 I_{\mathrm{s}}+B$ \\
$z$ & $N+M_z\pm I_z+\tfrac13 I_{\mathrm{s}}+B$ &
$M_y+\tfrac23 I_{\mathrm{s}}+B$ \\
\bottomrule
\end{tabular}
\caption{Longitudinal polarized-neutron cross sections for ideal polarization. The subscript $\alpha=x,y,z$ specifies the polarization axis; the first and second superscripts denote the incident and scattered spin states, with $+$ ($-$) denoting spin parallel (antiparallel) to
$\alpha$. $C_x$ denotes the chiral magnetic contribution, while $I_y$ and $I_z$ denote nuclear--magnetic interference contributions; all three change sign when the incident neutron polarization is reversed~\cite{101103PhysRev181920}. $N$ denotes the non-spin-flip nuclear scattering, $I_{\mathrm{s}}$ is nuclear spin-incoherent scattering, and $B$ represents additional instrumental and sample-environment background.}
\label{tab:crosssections}
\end{table}

The same LPA selection rules apply to quasielastic and inelastic scattering, where $M_y$ and $M_z$ represent the corresponding components of the dynamical magnetic correlation function rather than squared static magnetic structure factors~\cite{Zaliznyak2005,ShiraneShapiroTranquada2002}. For noncollinear structures, multiple magnetic sublattices, or more complex spin-wave spectra, the polarization dependence of the neutron cross section can become less straightforward to calculate. Established software packages can assist with both elastic and inelastic problems: Mag2Pol provides tools for polarized magnetic diffraction, while SpinW and Sunny can calculate polarization-resolved spin correlations and neutron-scattering cross sections for comparison with polarized inelastic measurements~\cite{Qureshi2019Mag2Pol,TothLake2015,Dahlbom2025Sunny}. In either case, the experiment first resolves the response in the local $y$ and $z$ directions. To connect these measured components to magnetic anisotropy in the crystal, one must relate the local polarization frame to the crystallographic axes.

\subsection{From local polarization to crystallographic components}
\label{sec:decomposition}

Polarization analysis resolves the magnetic scattering in a local frame tied to the measured wave vector $\Q$, giving the local responses $M_y(\Q)$ and $M_z(\Q)$. Magnetic anisotropy, however, is usually discussed in terms of the intrinsic response along the crystallographic axes, $M_a$, $M_b$, and $M_c$. Here we take $M_a$, $M_b$, and $M_c$ to denote the intrinsic crystallographic responses at the reduced wave vector $\bm{q}$ and energy $E$, with the magnetic form-factor dependence removed. The corresponding wave-vector dependence of the measured intensity is included in $A(\Q)$ below.

As a representative example, consider an orthogonal crystallographic basis $(a,b,c)$ with the sample aligned in the $(H,0,L)$ scattering plane. The local $z$ axis is parallel to $b$. The local $y$ axis lies in the $ac$ plane and changes its orientation as $\Q$ moves from one reciprocal-space zone to another [Fig.~\ref{fig:lpa_geometry}(b)]. When the off-diagonal $a$--$c$ correlation vanishes\footnote{In general, the $y$ projection can also contain a real $a$--$c$ cross-correlation term, $-2\operatorname{Re}M_{ac}\sin\theta\cos\theta$. Whether this term survives depends on symmetry and domain populations. For example, for $\bm{q}=(0,0,q)$, a preserved twofold rotation about $c$ forces it to vanish because the in-plane component changes sign while the $c$-axis component does not. This assumption requires particular care for elastic diffraction, where static noncollinear or canted components can interfere coherently and may themselves break the symmetry required for this term to vanish. For inelastic scattering, the equilibrium state often retains symmetries that forbid the corresponding off-diagonal dynamical correlation, while domain averaging can provide an additional cancellation. Thus, neglecting this term is often justified for magnetic fluctuations, although finite energy transfer alone does not require it to vanish.},
the measured local responses can be written as
\begin{equation}
M_y(\Q)=A(\Q)
\left(
M_a\cos^2\theta
+
M_c\sin^2\theta
\right),
\qquad
M_z(\Q)=A(\Q)M_b,
\label{eq:diagonal}
\end{equation}
where $\theta$ specifies the orientation of $\Q$ within the $ac$ plane, as defined in Fig.~\ref{fig:lpa_geometry}(b). The factor $A(\Q)$ collects the common wave-vector-dependent scale factors, including the magnetic form factor, structure-factor contributions, sample-illumination volume, instrument resolution, and normalization.

While $M_z$ directly measures $M_b$, $M_y$ at a single wave vector provides only one linear combination of $M_a$ and $M_c$. Separating the two therefore requires measuring the same intrinsic magnetic response at a second equivalent wave vector in a different reciprocal-space zone, where the polarization factors are different. For two such wave vectors, $\Q_1$ and $\Q_2$, the intrinsic quantities $M_a$, $M_b$, and $M_c$ are the same because the reduced wave vector $\bm{q}$ and energy transfer are the same; only the common scale $A(\Q)$ and the geometrical projections change.

Since $M_z(\Q)=A(\Q)M_b$, the relative scale of the $\Q_2$ measurement with respect to $\Q_1$ follows directly from \cite{PhysRevB.90.140502}
\begin{equation}
r\equiv
\frac{A(\Q_2)}{A(\Q_1)}
=
\frac{M_z(\Q_2)}{M_z(\Q_1)}.
\label{eq:relative_normalization}
\end{equation}
Once $r$ is known, the $\Q_2$ response can be placed on the intensity scale of $\Q_1$. Since the remaining common intensity scale does not affect the spin anisotropy, we omit it from the notation below. The two $y$ projections satisfy
\begin{equation}
\begin{pmatrix}
M_y(\Q_1)\\[3pt]
M_y(\Q_2)/r
\end{pmatrix}
=
\begin{pmatrix}
\cos^2\theta_1 & \sin^2\theta_1\\
\cos^2\theta_2 & \sin^2\theta_2
\end{pmatrix}
\begin{pmatrix}
M_a\\
M_c
\end{pmatrix}.
\label{eq:twoq_projection}
\end{equation}
Together, these two equations determine $M_a$ and $M_c$, while $M_b$ is obtained directly from $M_z$. In this way, measurements in two equivalent reciprocal-space zones can recover all three crystallographic spin components from the local polarization responses. For energy-dependent measurements, the inter-zone scale factor $r$ should be energy independent. It is therefore important to evaluate $r(E)=M_z(\Q_2,E)/M_z(\Q_1,E)$ across the measured energy range. If $r(E)$ is constant within experimental uncertainty, an average value of $r$ can be used to reduce statistical fluctuations. A systematic energy dependence indicates that the assumed inter-zone scaling is not valid and should be resolved before extracting the crystallographic magnetic components.

When selecting the two wave vectors, three considerations are particularly important. First, the symmetry and domain population at the relevant reduced wave vector must justify neglecting the off-diagonal cross term. Second, the measured $M_z$ response, corresponding to $M_b$ in the present geometry, must have sufficient statistical precision at both wave vectors to determine the relative scale $r$ reliably. Third, the polarization factors should differ sufficiently to allow a stable separation of $M_a$ and $M_c$. The determinant of the projection matrix is
$\cos^2\theta_1-\cos^2\theta_2$,
so nearly identical projection angles make the decomposition poorly conditioned and amplify the experimental uncertainty.

\subsection{Finite polarization and flipping-ratio corrections}
\label{sec:finite_flipping_ratio}

While polarization analysis is a powerful technique, in practice its accuracy is limited by the finite polarization efficiency of a real neutron beam. Polarized neutron beams are produced and analyzed using devices such as Heusler crystals, polarizing supermirrors, or polarized $^3$He spin filters. Spin flippers, which reverse the neutron spin by $180^\circ$, and magnetic coils are used to prepare the neutron polarization along the desired local $x$, $y$, or $z$ direction, while a weak guide field, typically of order $\sim$10-15 Gauss, maintains the spin polarization along the beam path. The exact implementation depends on the instrument, but none of these elements has perfect efficiency. A nominally pure spin state therefore contains a small admixture of the opposite state, leading to mixing between the nominal SF and NSF channels. This finite-polarization mixing appears experimentally as intensity leakage from one channel into the other. 

This leakage is characterized experimentally by the flipping ratio
\begin{equation}
R_\alpha=
\frac{I_{\alpha}^{\NSF}}
     {I_{\alpha}^{\SF}},
\label{eq:FR_definition}
\end{equation}
where the intensities are measured on a suitable nuclear reflection with negligible intrinsic SF scattering, after background subtraction~\cite{101103RevModPhys87855}. In many experiments, a representative value is determined before the measurement and a common flipping ratio $R$ is then used for the three polarization directions and nearby wave vectors. Although the flipping ratio $R_\alpha$ can in principle be different for $x,y,z$ channels, the differences are usually small and do not materially affect the experimental conclusions. Therefore, a common polarization-independent $R$ is often an adequate approximation, but greater care is needed when the intrinsic magnetic signal is weak or when a large nuclear or background contribution can leak into the measured channel.

This leads to an important practical point: a finite signal in a SF channel of $x\parallel\Q$ is not, by itself, evidence for magnetic scattering. With finite polarization, strong nuclear scattering at the same wave vector can leak into the nominal SF channel and produce an apparent signal. The relevant observable is therefore the polarization-dependent contrast after this leakage has been accounted for.

To illustrate the correction of finite polarization, we assume a common flipping ratio $R$ for $x$, $y$, and $z$, symmetric leakage between SF and NSF channels, and a common polarization-independent background. We further assume that chiral and nuclear--magnetic interference terms are absent or have been removed by averaging the corresponding measurements with reversed incident spin states. Under these assumptions, the six measured channels are related to these ideal contributions by~\cite{101103RevModPhys87855}
\begin{equation}
\begingroup
\renewcommand{\arraystretch}{1.3}
\begin{pmatrix}
\sigma_x^{\SF}\\
\sigma_y^{\SF}\\
\sigma_z^{\SF}\\
\sigma_x^{\NSF}\\
\sigma_y^{\NSF}\\
\sigma_z^{\NSF}
\end{pmatrix}_{\!\mathrm{meas}}
=
\frac{1}{R+1}
\begin{pmatrix}
R & R & 1 & (2R+1)/3 & R+1\\
1 & R & 1 & (2R+1)/3 & R+1\\
R & 1 & 1 & (2R+1)/3 & R+1\\
1 & 1 & R & (R+2)/3 & R+1\\
R & 1 & R & (R+2)/3 & R+1\\
1 & R & R & (R+2)/3 & R+1
\end{pmatrix}
\begin{pmatrix}
M_y\\
M_z\\
N\\
I_{\mathrm{s}}\\
B
\end{pmatrix}.
\endgroup
\label{eq:flippingmatrix}
\end{equation}

Here, $M_y$ and $M_z$ are the magnetic scattering contributions defined above, $N$ denotes the nuclear scattering, $I_{\mathrm{s}}$ the nuclear spin-incoherent scattering, and $B$ the polarization-independent background. The magnetic components can then be obtained directly from differences between the measured polarization channels. For $\alpha=y,z$,
\begin{equation}
M_\alpha=
\frac{R+1}{R-1}
\left(\sigma_x^{\SF}-\sigma_\alpha^{\SF}\right)_{\mathrm{meas}}
=
\frac{R+1}{R-1}
\left(\sigma_\alpha^{\NSF}-\sigma_x^{\NSF}\right)_{\mathrm{meas}}.
\label{eq:contrasts}
\end{equation}
Thus, the common nonmagnetic contributions cancel in the channel differences, while the factor $(R+1)/(R-1)$ restores the magnetic contrast reduced by imperfect polarization. Agreement between the SF- and NSF-derived values provides a useful internal consistency check of the measurements. As $R$ approaches 1, corresponding to the unpolarized limit, the inversion becomes ill-conditioned.

In many experiments, it is adequate to assume the same flipping ratio for the three polarization directions when extracting $M_y$ and $M_z$. Extra care is required, however, when the magnetic signal of interest is small compared with the nonmagnetic scattering, including nuclear coherent, nuclear spin-incoherent, and background contributions. The flipping ratio is not a universal constant of the instrument and can depend on polarization direction, wave vector, temperature, sample environment, and details of the polarization setup~\cite{101103PhysRevB110195109,Croft2017FlippingRatio,Bourges2019TechnicalAssessment}. A small difference between $R_x$ and $R_y$, for example, produces unequal leakage of a large nuclear intensity into $\sigma_x^{\SF}$ and $\sigma_y^{\SF}$, so their difference can mimic a weak magnetic signal. Similarly, an instrumental asymmetry between the $-+$ and $+-$ channels for $x$ polarization can produce an apparent chiral signal. Uncertainty in the flipping-ratio correction can therefore be as important as counting statistics in weak-signal measurements.

The debate over proposed intra-unit-cell magnetism in YBa$_2$Cu$_3$O$_{6+x}$ provides a concrete example. Croft \emph{et al.}\ showed that small changes in sample orientation and lattice parameters can sample different regions of the polarizing optics and alter the effective polarization~\cite{Croft2017FlippingRatio}. Bourges \emph{et al.}\ emphasized beam-polarization inhomogeneity, sample displacement with temperature, and the resulting uncertainty in the leakage baseline~\cite{Bourges2019TechnicalAssessment}. Their differing interpretations illustrate why weak-signal measurements require the stability of this baseline to be established and its uncertainty to be included alongside counting statistics.

When weak magnetism is measured on top of a strong nuclear reflection, three practical questions are particularly important:
\begin{enumerate}
    \item Is the flipping ratio sufficiently well characterized at this $\Q$, or under representative nearby conditions?
    \item Are $R_x$, $R_y$, and $R_z$ equal within the accuracy required by the claimed magnetic signal?
    \item Does the magnetic contrast remain significant after the uncertainty and possible systematic variation of the flipping ratio are propagated into the result?
\end{enumerate}

Nearby nuclear reflections and checks of the polarization-direction dependence at representative experimental conditions can help to establish whether the assumed flipping ratio is appropriate. When the uncertainty in the leakage correction becomes comparable to the magnetic contrast, it should be included as a systematic uncertainty in the extracted $M_y$ and $M_z$. 

The practical lesson is simple: polarization efficiency should be treated as part of the measurement rather than as a fixed instrument constant. Before assigning a small difference between two channels to magnetism, one should first ask whether a plausible difference in leakage could produce the same effect.
The discussion so far has focused on how LPA is analyzed. We now turn to how it is used in practice, through several representative experiments.

\section{Identifying weak magnetic scattering}
\label{sec:background}

In most unpolarized neutron scattering experiments, temperature-difference spectra between low (possible magnetic signal) and high (featureless background) temperatures are commonly used to estimate the magnetic scattering. However, near zero energy transfer, an apparent absence of positive scattering can reflect the increased background scattering at high temperature instead of the lack of magnetic signal at low temperature. This is a particular concern in rare-earth magnets, where nonmagnetic quasielastic scattering can overwhelm the low-energy magnetic response. Ce$_2$Zr$_2$O$_7$, a candidate for quantum spin ice, makes the distinction consequential: its near-zero-energy spectrum is used to test predicted gapless spin excitations from emergent photons~\cite{101038s41567025029229}.

\begin{figure}[htbp]
\centering
\includegraphics[width=0.85\textwidth]{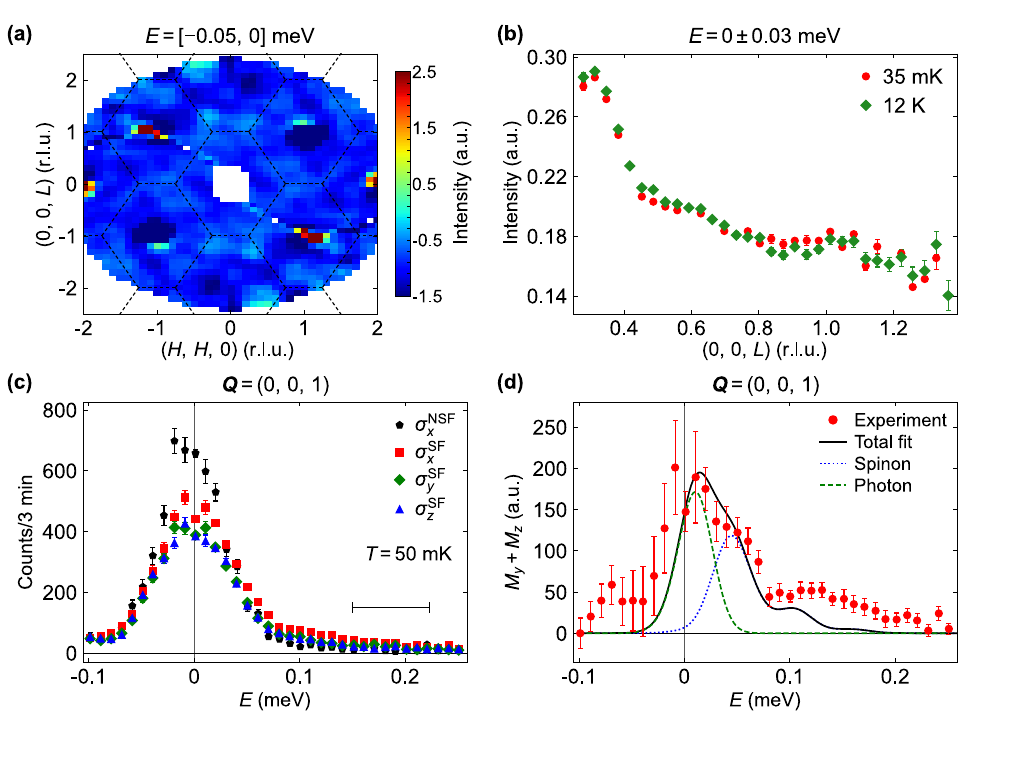}
\small
\caption{\textbf{Background separation near zero energy in Ce$_2$Zr$_2$O$_7$.}
\textbf{(a)} Unpolarized intensity at 35~mK after subtraction of the 12~K reference, integrated over $-0.05\leq E\leq0$~meV. The negative intensity illustrates oversubtraction near the elastic line. Dashed lines mark Brillouin-zone boundaries.
\textbf{(b)} Unpolarized intensity along $(0,0,L)$ at $E=(0\pm0.03)$~meV, measured at 35~mK (red circles) and 12~K (green diamonds). The higher reference-temperature intensity over much of the scan illustrates the origin of the oversubtraction.
\textbf{(c)} Measured $\sigma_x^{\NSF}$ and $\sigma_{x,y,z}^{\SF}$ energy scans at $\Q=(0,0,1)$ and 50~mK, with final neutron energy 3.23~meV. The horizontal bar indicates the energy-resolution FWHM.
\textbf{(d)} The magnetic response $M_y+M_z$ extracted from the polarization channels in (c). The polarization channels allow the magnetic response to be separated at the measurement temperature without using a high-temperature reference. Black solid, blue dotted and green dashed curves are the published total fit, spinon contribution and photon contribution, respectively.
Vertical error bars denote one standard deviation in (b,c) and propagated uncertainties in (d).
Unpolarized measurements in (a,b) were performed on the Cold Neutron Chopper Spectrometer (CNCS) at the Spallation Neutron Source, Oak Ridge National Laboratory; polarized measurements in (c,d) were performed on the ThALES spectrometer at the Institut Laue Langevin.
Panel (a) is replotted from Fig.~3d of Gao \textit{et al.}~\cite{Gao2019Ce}, retaining its original color scale. Data and curves in (b--d) are digitized and replotted from Extended Data Fig.~4c and Figs.~2a, 2d and 2g of Gao \textit{et al.}~\cite{101038s41567025029229}.}
\label{fig:ce227_background}
\end{figure}

In the 2019 unpolarized study, Gao and co-workers estimated the magnetic response by subtracting the 12~K data from the 35~mK data (Fig.~\ref{fig:ce227_background}a). Magnetic scattering at 12 K was treated as sufficiently diffuse and independent of wave vector and energy to be included in the background~\cite{Gao2019Ce}. At fixed $(\Q,E)$, writing $I=M+B_{\mathrm{nm}}$ with $M=M_y+M_z$ gives
\begin{equation}
I(T_{\mathrm{low}})-I(T_{\mathrm{high}})
=
\bigl[M(T_{\mathrm{low}})-M(T_{\mathrm{high}})\bigr]
+
\bigl[B_{\mathrm{nm}}(T_{\mathrm{low}})
-B_{\mathrm{nm}}(T_{\mathrm{high}})\bigr].
\label{eq:temperature}
\end{equation}
The difference is therefore a change in neutron scattering spectra on cooling, not automatically the low-temperature magnetic spectrum. Quantitative recovery requires the reference magnetic signal and the temperature-dependent background to be negligible or corrected for. Incoherent scattering, instrumental and sample-environment contributions, and low-energy phonons need not share the same temperature dependence. A single Bose-factor correction cannot generally describe all of them. The lack of positive difference intensity near zero energy was consistent with a gap interpretation, but the subtraction alone could not establish such a gap \cite{Gao2019Ce}. 

The polarized experiment addressed this ambiguity at a single temperature (Fig.~\ref{fig:ce227_background}c,d). For ideal or polarization-corrected channels, with the chirality absent or averaged out, $2\sigma_x^{\SF}-\sigma_y^{\SF}-\sigma_z^{\SF}=M_y+M_z$ cancels the common nonmagnetic contributions. Under the common-$R$ assumptions of Sec. 2.3, the same combination of measured channels must be multiplied by $(R+1)/(R-1)$. Particular care is required when substantial nonmagnetic scattering from the sample or sample environment is present, as can occur in dilution-refrigerator measurements, because even a small polarization dependence of $R_\alpha$ can generate an apparent magnetic contrast, as discussed in Sec. 2.3. Using this method, Gao and co-workers recovered near-zero-energy magnetic weight missed in the earlier unpolarized neutron scattering experiments and attributed its apparent absence to an enhanced near-zero-energy background at 12 K~\cite{101038s41567025029229}. The recovered spectrum, together with thermodynamic measurements, was compared with a quantum spin ice description involving emergent photons and fractionalized excitations. Polarization analysis changed the interpretation by separating a weak magnetic response from nonmagnetic scattering without relying on a second temperature as its background.

The lesson extends beyond Ce$_2$Zr$_2$O$_7$. In quantum spin-liquid candidates, the structure of the lowest-energy spectrum is often central to distinguishing competing ground states, so an error in background subtraction can change the physical interpretation. For example, the continuum in NaYbSe$_2$ was interpreted in terms of a spinon Fermi surface~\cite{Dai2021NaYbSe2}, whereas a gap near $\Gamma$ together with gapless spectral weight at $K$ and $M$ supported a Dirac spin-liquid interpretation in YbZn$_2$GaO$_5$~\cite{Bag2024YbZn2GaO5}. Calculations for the triangular-lattice $J_1$--$J_2$ model further show why the lowest-energy response matters for this distinction~\cite{Sherman2023TriangularSpectrum}. Similar polarization analyses are therefore desirable for these candidates, since resolving such momentum-dependent gaps requires not only sufficient energy resolution but also reliable subtraction of the nonmagnetic background.

\section{Resolving weak moment components}
\label{sec:constraints}
A small additional moment component can change the interpretation of a material's electronic properties, yet is difficult to establish when its scattering overlaps with a strong nuclear reflection. LPA can detect that component or place a model-dependent upper bound on it. The relevant evidence is the corrected channel contrast, not the existence of an SF peak alone.

In YbMnBi$_2$, the giant anomalous Nernst effect and large anomalous Hall response have been attributed to a Weyl semimetal state induced by canting of the Mn moments~\cite{Pan2022YbMnBi2ANE,Xie2026}. This scenario requires an in-plane component of the predominantly $c$-axis antiferromagnetic Mn moments. The magnetic structure therefore provides a direct test of that specific scenario.

\begin{figure}[htb]
\centering
\includegraphics[width=0.85\textwidth]{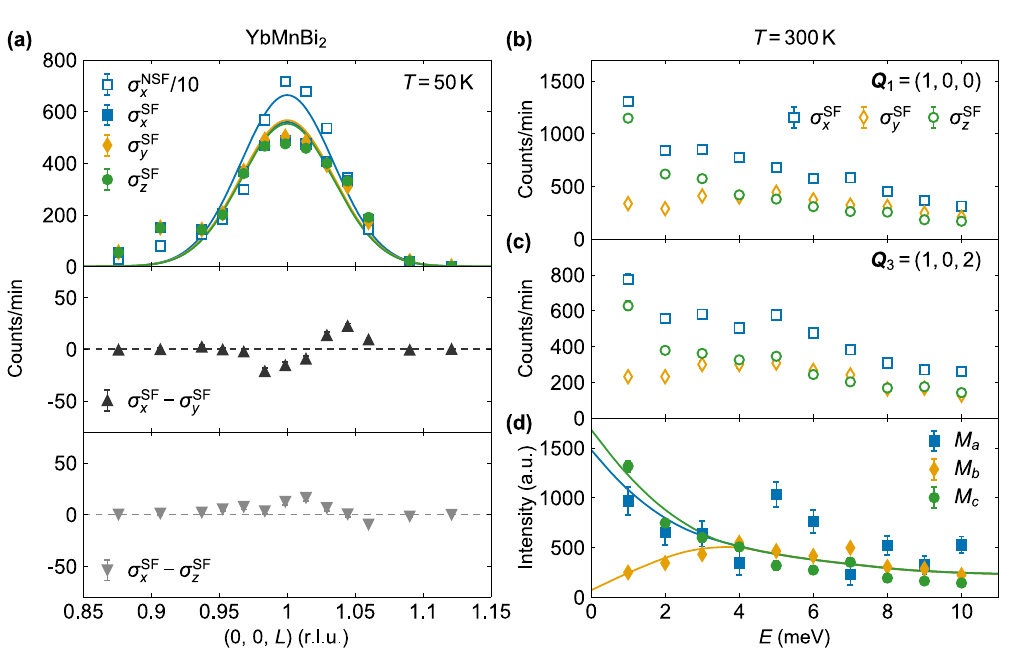}
\small
\caption{\textbf{Polarized-neutron scattering measurements of YbMnBi$_2$.}
\textbf{(a)} Elastic scan along $(0,0,L)$ at $T=50$~K, showing the non-spin-flip channel $\sigma_x^{\NSF}/10$ and the three spin-flip channels $\sigma_x^{\SF}$, $\sigma_y^{\SF}$, and $\sigma_z^{\SF}$. The two lower plots show $\sigma_x^{\SF}-\sigma_y^{\SF}$ and $\sigma_x^{\SF}-\sigma_z^{\SF}$; dashed lines mark zero.
\textbf{(b,c)} Spin-flip energy scans at $\Q_1=(1,0,0)$ and $\Q_3=(1,0,2)$, respectively, at $T=300$~K, following the notation of Ref.~\cite{Xie2026}.
\textbf{(d)} Extracted magnetic response components $M_a$, $M_b$, and $M_c$ along the crystallographic axes at $T=300$~K. Solid curves reproduce the Gaussian profiles in (a) and the guides to the eye in (d) from the source figures. Error bars show the supplied uncertainties; difference-channel uncertainties are propagated in quadrature assuming independent errors. Replotted from the numerical data corresponding to Figs.~2(d), 3(b), 3(f), and 3(j) of Xie \textit{et al.}~\cite{Xie2026}.}
\label{fig:ymb_polarization}
\end{figure}

Xie, Xu, and co-workers searched at $\Q=(0,0,1)$, where the scattering geometry excludes the dominant $c$-axis moment while retaining sensitivity to a uniform in-plane component~\cite{Xie2026}. At 50~K, a peak was visible in all three SF channels (Fig.~\ref{fig:ymb_polarization}a). Crucially, the peak intensities matched: the differences $\sigma_x^{\SF}-\sigma_y^{\SF}$ and $\sigma_x^{\SF}-\sigma_z^{\SF}$ contained no resolved magnetic signal within their common-$R$ analysis. The peaks were therefore assigned to nuclear leakage. Comparing the residual intensity with the $c$-axis magnetic intensity at $(1,0,0)$, with the appropriate structure factors and geometrical corrections, placed an upper limit of approximately $0.3^{\circ}$ on uniform canting. This result contrasts with the proposed canting-induced bulk Weyl scenario, where a much larger canting angle 
is expected \cite{Pan2022YbMnBi2ANE}.

\begin{figure}[htbp]
\centering
\includegraphics[width=0.85\textwidth]{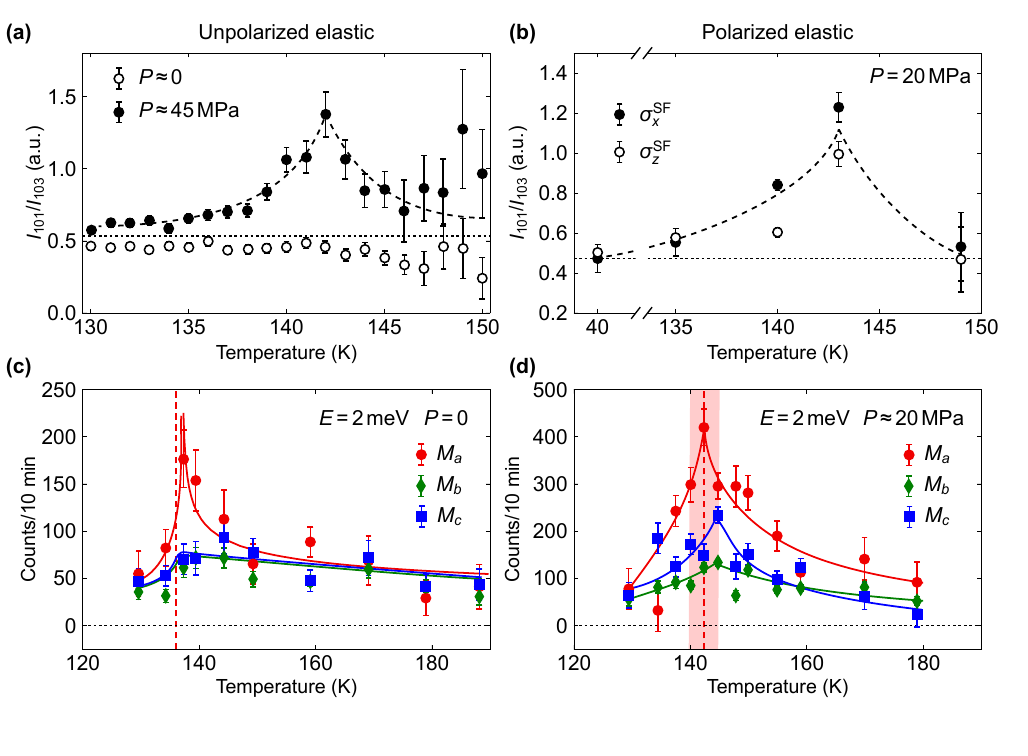}
\small
\caption{\textbf{Elastic and critical spin responses in BaFe$_2$As$_2$ under uniaxial pressure.}
\textbf{(a)} Temperature dependence of the ratio $I_{101}/I_{103}$ of integrated elastic intensities at $(1,0,1)$ and $(1,0,3)$, measured with unpolarized neutrons at approximately 0~MPa (open black circles) and 45~MPa (filled black circles).
\textbf{(b)} The same ratio under 20~MPa, obtained from polarized $\sigma_x^{\mathrm{SF}}$ (filled black circles) and $\sigma_z^{\mathrm{SF}}$ (open black circles). Its increase near the transition is consistent with an additional static or quasistatic $c$-axis component. The broken temperature axis is retained.
\textbf{(c,d)} Temperature dependence of the extracted spin-fluctuation components $M_a$ (red circles), $M_b$ (green diamonds) and $M_c$ (blue squares) along the crystallographic axes at $E=2$~meV, without and with approximately 20~MPa, respectively. The pressure-induced critical enhancement of $M_c$, without a comparable enhancement of $M_b$, reveals the change in spin anisotropy. The vertical count scales differ between (c) and (d), and absolute count levels are not directly comparable between the two experiments. Vertical transition markers, shading and guide curves are retained from the source; error bars represent one standard deviation. Data and curves digitized and replotted from Figs.~4(e), 4(d), 3(g) and 3(h) of Liu \textit{et al.}~\cite{101038s41467020194215}.}
\label{fig:bfa_pressure}
\end{figure}

In BaFe$_2$As$_2$, a prototypical parent compound of the 122 iron-based superconductors, uniaxial compressive pressure is routinely applied to induce lattice strain and select a single orthorhombic domain \cite{101103RevModPhys87855}. The Fe moments normally lie along the $a$-axis direction, while the uniaxial strain used for detwinning along the $b$-axis direction can modify this arrangement. Liu and co-workers examined the integrated elastic $\sigma_z^{\SF}$ intensities at $(1,0,1)$ and $(1,0,3)$, which are more sensitive to the $c$- and $a$-axis components, respectively~\cite{101038s41467020194215}. A change in the magnitude of an $a$-axis moment would scale both peaks together. Instead, $I_{101}/I_{103}$ increased near $T_\mathrm{N}$ under uniaxial pressure (Fig.~\ref{fig:bfa_pressure}b); complementary unpolarized measurements found no such enhancement without pressure. The authors interpreted this change as a static or quasistatic $c$-axis antiferromagnetic component. It could represent a uniform tilt toward $c$ or long-lived domains with different moment directions in the $ac$ plane. In either interpretation, detwinning changes the magnetic correlations near the transition, with consequences that also appear in the low-energy excitations. Here ``quasistatic'' is relative to the experimental energy resolution.

The two studies give complementary outcomes: one constrains a proposed uniform canting, while the other detects an additional magnetic component induced by strain. In each case, the information comes from choosing projections and polarization channels that distinguish the weak component from the dominant scattering. While these examples focus on static magnetic order, the same materials also illustrate the complementary role of polarization analysis in probing magnetic dynamics, where it can resolve the spin-space character of magnetic excitations in addition to constraining the orientation of the ordered moment, as discussed in the next section.

\section{Spin-space anisotropy of magnetic excitations}
\label{sec:anisotropy}

Unpolarized inelastic neutron scattering measures the total magnetic response perpendicular to $\Q$, but does not generally determine the polarization of the fluctuating moments. LPA can separate these spin-space components and thereby reveal anisotropy that is hidden in the total magnetic intensity. Following the procedure introduced in Sec.~\ref{sec:decomposition}, measurements at appropriately chosen wave vectors can resolve the magnitude of spin fluctuations along the crystallographic axes~\cite{101103RevModPhys87855}.

\subsection{\texorpdfstring{YbMnBi$_2$}{YbMnBi2}: anisotropic fluctuations and anomalous transport}

In YbMnBi$_2$, the anomalous Hall and Nernst responses remain appreciable in the paramagnetic state near $T_{\mathrm{N}}$ but become much smaller both deep in the magnetically ordered and in the high-temperature paramagnetic states ~\cite{Pan2022YbMnBi2ANE,Xie2026}. At the same time, polarized neutron diffraction places a stringent upper limit on uniform spin canting, as discussed in Sec.~\ref{sec:constraints}, motivating a closer examination of spin fluctuations in the paramagnetic state near $T_{\mathrm{N}}$.

Polarized inelastic measurements at equivalent antiferromagnetic wave vectors $(1,0,0)$ and $(1,0,2)$ were used to separate the three crystallographic components (Fig.~\ref{fig:ymb_polarization}b,c). At 300~K, the low-energy response below approximately 4~meV satisfies $M_a \simeq M_c > M_b$. At higher energies, this anisotropy is reduced and the three components become more comparable, approaching a more isotropic response around 9 meV (Fig.~\ref{fig:ymb_polarization}d)~\cite{Xie2026}.

All three components are enhanced near $T_\mathrm{N}$, while the in-plane anisotropy $M_a-M_b$ also grows strongly near the transition. This behavior cannot be understood solely as a conventional enhancement of longitudinal $c$-axis critical fluctuations.

Isostructural CaMnBi$_2$ exhibits a much smaller anomalous Hall response and no comparable in-plane spin-space anisotropy~\cite{Xie2026}. This contrast supports the interpretation that paramagnetic excitations in YbMnBi$_2$ form a dynamic spin-nematic phase associated with spin-orbit coupling. 
The key observation is that $M_a\neq M_b$ can be established even when the corresponding fluctuations overlap in momentum and energy, revealing directional spin correlations that would be impossible to identify from the unpolarized intensity alone.

A spin-nematic response, reflecting anisotropy in dynamic two-spin correlations, need not by itself break time-reversal symmetry. Connecting these anisotropic fluctuations to anomalous transport requires an additional time-reversal-breaking ingredient. Xie \textit{et al.} proposed that an in-plane magnetic field couples to the Yb$^{3+}$ moments and induces dynamic scalar spin chirality, providing a possible link between the fluctuating magnetic state and the observed anomalous Hall and Nernst effects~\cite{Xie2026}. This three-spin scalar chirality is not directly measured by neutron scattering.

\subsection{Iron pnictides: uniaxial-strain-induced spin anisotropy}

In-plane uniaxial strain is routinely used to detwin BaFe$_2$As$_2$ and select a single magnetic domain \cite{101103RevModPhys87855}. Whether it also changes the spin correlations is important for interpreting measurements under strain. If the uniaxial strain acted only as a domain-selection field, the low-energy critical fluctuations near $T_\mathrm{N}$ would be expected to retain their predominantly $a$-axis polarization as in the unstrained case \cite{Wilson2010CriticalFluctuations}.

Liu and co-workers found that this simple picture does not hold 
for BaFe$_2$As$_2$ under uniaxial pressure \cite{101038s41467020194215}. For fluctuations polarized purely along the $a$ axis, the relative magnetic intensities at $(1,0,1)$ and $(1,0,3)$ are fixed by their different polarization factors, so their intensity ratio should remain approximately unchanged as the critical scattering develops. Experimentally, however, the relative intensity between these two wave vectors changed strongly near $T_\mathrm{N}$, indicating the emergence of an additional spin component.

Polarization analysis identified this additional response as predominantly $c$-axis polarized. Measurements of the three SF channels at $(1,0,1)$ and $(1,0,3)$ are particularly useful because the two wave vectors weight $M_a$ and $M_c$ differently, while $M_z=M_b$ in the $(H,0,L)$ scattering plane. Combining the two measurements using the decomposition of Sec.~\ref{sec:decomposition} showed a pronounced enhancement of $M_c$ near the magnetic transition, with no comparable increase in $M_b$ (Fig.~\ref{fig:bfa_pressure}c,d)~\cite{101038s41467020194215}.

The appearance of $c$-axis critical fluctuations indicated that the applied strain was modifying the magnetic anisotropy rather than simply changing the magnetic-domain population. Guided by this result, polarized neutron diffraction revealed a corresponding static or quasistatic $c$-axis antiferromagnetic component near $T_\mathrm{N}$, as discussed in Sec.~\ref{sec:constraints}~\cite{101038s41467020194215}. The authors attributed the coupling between the in-plane lattice distortion and the out-of-plane magnetic response to spin-orbit coupling.

The intensity ratio first signaled a change; the polarization channels identified the component responsible. In BaFe$_2$As$_2$, resolving $M_a$, $M_b$, and $M_c$ showed that the uniaxial strain routinely used for detwinning also modifies the magnetic correlations themselves. This distinction is important when interpreting neutron, transport, and spectroscopic measurements performed under uniaxial strain.

\section{Outlook}

% As increasingly subtle magnetic states are recognized in quantum materials, measurements of magnetic intensity alone are often insufficient to distinguish competing microscopic pictures. Information about spin polarization, chirality and magnetic-domain populations is becoming increasingly important. These are among the quantities for which neutron polarization analysis provides information beyond conventional unpolarized scattering.

Polarized neutron scattering is becoming increasingly important as a tool to unveil the magnetic properties of quantum materials.  Using LPA, as discussed here, and spherical neutron polarimetry with a Cryopad \cite{TASSET199969}, one can obtain complementary, momentum-resolved information about magnetic structures and dynamics. Their particular strength is to separate spin-dependent contributions that may overlap in unpolarized neutron scattering experiments.

One timely direction beyond the examples discussed here is altermagnetism. In an altermagnet, opposite N\'eel-vector domains carry opposite altermagnetic order, so spin-polarized responses can cancel when the time-reversal-related domains are equally populated. Their relative populations, however, can be biased by intrinsic or extrinsic strain~\cite{Liu2026MnTeStrain} and by appropriate magnetic field-cooling protocols~\cite{Amin2024MnTeDomains,Aoyama2024MnTePiezomagnetism}, making the domain population itself an important quantity to determine. Unpolarized neutron diffraction does not distinguish an exact time-reversed pair through the squared magnetic structure factor. Polarized diffraction can constrain their relative populations through nuclear-magnetic interference at reflections with suitable nonzero nuclear and magnetic amplitudes \cite{McClarty2025}. The allowed interference depends on symmetry and geometry; it is not a universal identifier of altermagnetism on its own.

The dynamical counterpart is magnon chirality. Polarized inelastic scattering can resolve the chirality of the split magnon branches, information that is likewise absent from the unpolarized intensity. McClarty, Gukasov and Rau identified these static and dynamical polarization signatures as characteristic probes of altermagnetism~\cite{McClarty2025}. Recent experiments on $\alpha$-MnTe and MnF$_2$ have begun to demonstrate these capabilities. In $\alpha$-MnTe, nuclear--magnetic interference was used to detect and control the N\'eel-vector domain population, while polarized inelastic scattering directly observed chiral magnons and their field-controlled reversal~\cite{Liu2026Domains,Liu2026Chirality}. In MnF$_2$, polarized inelastic measurements found opposite chiral responses of the split magnon modes, while polarized diffraction reconstructed the magnetization density and resolved anisotropic magnetic contributions associated with the altermagnetic state~\cite{Faure2025,Kibalin2026}.

Similar approaches could address open questions in unconventional superconductors and kagome magnets. In UTe$_2$, earlier inelastic neutron scattering established antiferromagnetic spin fluctuations and a magnetic resonance associated with superconductivity~\cite{101038s41586021041515}. More recently, polarized inelastic measurements resolved their spin-space anisotropy at the $Y$ point, finding an approximately isotropic response within the $ac$ plane, $M_a\simeq M_c$, with no observable $b$-axis component~\cite{Wang2026UTe2}. This result challenges the previously assumed $a$-axis Ising response and places direct constraints on the possible superconducting pairing channels. In the kagome metal FeGe, polarized neutron scattering could determine whether the incommensurate magnetic order and its low-energy excitations are purely transverse or contain a longitudinal component, providing a direct test of the microscopic origin and constraining competing descriptions of the magnetic instability~\cite{101038s41467023441902,Klemm2025Vacancy}.

The main practical cost of polarization analysis is reduced count rate. Polarizing and analyzing the neutron beam reduces intensity, making measurements of small samples and weak signals demanding. Finite polarization, domain averaging, sample depolarization, and restrictions imposed by sample environments also require care. Quantitative interpretation therefore requires appropriate polarization corrections together with careful consideration of the experimental geometry and the magnetic cross terms allowed by symmetry.

The examples in this review cover only a fraction of the problems that benefit from neutron polarization analysis. Larger samples, brighter sources, improved polarization methods, and more versatile sample environments will extend its use in determining magnetic ground states and spin dynamics in quantum materials.

\section*{Acknowledgements}
We thank Yaofeng Xie, Steven J. Gomez Alvarado, and Tong Chen for helpful comments and discussions. The neutron scattering work at Rice is supported by the U.S. DOE BES DE-SC0026179 (P.D.). Part of the materials work at Rice is supported by the Robert A. Welch Foundation under Grant No. C-1839 (P.D.).

\section*{Declaration of Competing Interest}
The authors declare that they have no competing interests.

\bibliographystyle{unsrtnat}
\bibliography{references}
\end{document}